\documentclass[12pt]{article}
\usepackage{amssymb,amsmath, psfrag, epsfig}

\def\ov{\overline}
\font\er = cmr8

\def\u{\vskip  .075 in}
\def\nh{\noindent\hangindent=1 true cm \hangafter = 1}

\def\u{\vskip  .1 in}

\def\ov{\overline}
\def\be{\begin{equation}}
\def\ee{\end{equation}} 

\begin{document}

\title{{\bf Predicting the  probability of persistence of HIV infection with
the standard model}\\
{\small {\bf  Henry C. Tuckwell$^{1\dagger}$, Patrick D. Shipman$^{2}$ }}\\   \
\  \\ 
{\small $^1$ Max Planck Institute for Mathematics in the Sciences\\
Inselstr. 22, 04103 Leipzig, Germany\\
$^2$ Department of Mathematics, University of Colorado\\
Fort Collins, CO 80523-1874, USA\\
$^{\dagger}$ {\it Corresponding author}: tuckwell@mis.mpg.de}}

\maketitle

\newpage

\begin{abstract}
We consider the standard three-component differential equation
model for the growth of an HIV virion population in an infected host in the
absence of drug therapy. 
The dynamical properties of the model are determined by the
set of values of six parameters which vary across host populations.
There may be one or  two critical points whose natures play a key role in
determining the outcome of infection and in particular whether
the HIV population will persist or become extinct. There are two cases which may  arise.
In the first case, there is only one critical point $P_1$ at biological values
and this is an asymptotically stable node. The system ends up with zero virions
and so the host becomes HIV-free. In the second case,  there are two critical
points $P_1$ and $P_2$ at biological values. Here $P_1$ is an unstable saddle
point and $P_2$ is an asymptotically stable spiral point with a non-zero virion
level.  In this case the HIV population persists unless parameters change. 
We let the parameter values take random values  from distributions based 
on empirical data, but suitably truncated,  and determine the 
probabilities of occurrence of the various combinations of critical
points. From these simulations the probability that an HIV
infection will persist, across a population, is estimated.  It is found that 
with conservatively estimated distributions of parameters, within the 
framework of the standard 3-component model, the chances  that
a within host HIV population will become extinct
is between 0.6\% and 6.9\%. With less conservative parameter estimates,
the probability is estimated to be as high as 24\%. 
The many factors related to the transmission and possible spontaneous elimination of the virus
are discussed.

\end{abstract}

\noindent {\it Short Title:} HIV persistence in a new host   

\noindent \it Keywords and Phrases: HIV infection, Model, Persistence 
\rm 
\section{Introduction and theory}
A comprehensive understanding and analysis of the early stages of HIV infection,
including the process of transmission, is important as it may lead to efficient
methods of reducing the probability that the virus successfully establishes 
itself in a new host. In a previous article  (Tuckwell  et al., 2008)
we have addressed the problem of
estimating the probability of a successful transmission of HIV infection to a new host.

There have been several mathematical models for the growth of within-host 
HIV populations, many of which are deterministic, for example Perelson et al. (1993), Phillips (1996)  and Perelson et al. (1996),
whereas others have
incorporated chance mechanisms (Tan and Wu, 1998; Tuckwell and LeCorfec, 1998;
Pearson et al., 2010). Such models have provided valuable 
insights into the time-course of viral dynamics and the
effects of drug therapy.  The simple 3-component differential equation model we examine in
this article has been successfully employed to predict the temporal evolution of HIV populations
in the primary stages of infection. As pointed out by Stafford et al. (2000),  although 
some  models have been hypothesized which describe the
progression to acquired immunodeficiency
syndrome  (AIDS) (for example, Essunger and Perelson, 1994) 
 useful information and possible prognoses can be obtained
from a knowledge of the viral dynamics in the  early stages.
In untreated HIV
infection, the risk of AIDS 
is known to be small until the 
CD4 cell count has reached low levels 
 or the viral load has reached high levels (Phillips et al., 2001).  
Indeed, our approach is to examine possible outcomes 
predicted by a model of primary infection for various
parameter values to ascertain whether the virus will 
be likely to persist after the initial infection period or possibly
 spontaneously die out due simply to dynamical system properties. 
 It is known from clinical studies that if HIV persists into and 
 beyond the primary stage then the baseline CD4 cell count is a 
 major predictor of eventual outcome under highly active
antiretroviral therapy (HAART)  (Egger et al., 2002).  Our analysis is
only applicable in the absence of HAART  which can  decrease 
the plasma viral load below the limit of detection  (Gallant et al., 2006).  
Extensions to include the effects of such therapy will be considered elsewhere. 

\u
The matter of extinction of an HIV
population in an individual new host also has been raised by some authors using stochastic
differential equation or similar models (Kamina et al., 2001; Merrill, 2005; Pearson et al., 2010). 
In this article, however, we are concerned with estimating the probability that an HIV population
does in fact become established in a host after a successful transmission on the assumption
that the viral population evolves deterministically according to a well-known dynamical
model.  The probability of persistence is determined by the distributions
of the parameters which describe the host's immune response.   Thus, as in accordance with
the standard deterministic model, extinction does not depend on the number of
virus particles which are become initially established in the new host.  
The veracity of such statements is contingent on the accuracy of the mathematical
model, which due to its simplicity is only expected to give an approximate prediction 
of outcomes.

\section{Model description}
Dynamical modeling of the growth of HIV populations within infected hosts is complicated
by spatial inhomogeneities, due for example to the occurrence of various reservoirs, such as those particularly rapidly
established in lymphatic tissues (Finzi and Siliciano, 1998; Pierson et al., 2000;  Pope and Haase, 2003; Kim and Perelson, 2006). 
However, and perhaps surprisingly, it seems that
the growth of HIV populations can be satisfactorily described even when such 
 inhomogeneities are  ignored,  which is the usual approach (Perelson, 2002)  and the one adopted here.
Some  models had additional  components representing
resting and latently infected cells (Perelson et al, 1993; Phillips, 1996) but
 a relatively successful  (Stafford et al., 2000)  now-accepted simple time-dependent  three-component model 
 for the evolution of within-host
HIV virion numbers in human or simian hosts, without any spatial variables,
has been employed for the last fifteen years or so.  This model (Nowak and Bangham, 1996; Bonhoeffer et al., 1997)
has as component variables,  at time $t$, 
 $T(t)$, the number of target or activated CD4$^+$ T-cells, 
$T^*(t)$, the number of productively infected such cells and $V(t)$ the number of free virus. 
In the early stages of HIV infection, to about 100 days, and in the absence of drug
therapy, these
quantities satisfy approximately the following three deterministic ordinary differential equations
\be \frac{dT}{dt}= \lambda - \mu T - kTV \ee
\be \frac{dT^*}{dt}= kTV-\delta T^* \ee
\be \frac {dV}{dt}= pT^* - cV. \ee  
In Table 1 are shown the variables  and 
parameters with their units. 

 In a recent article on simian immunodeficiency infection (Vaidya et al, 2010), an infection rate $k=k(t)$ which decreases
exponentially in time to an asymptotic value has been found to give a better fit for the growth 
of the viral population. This aspect can be easily incorporated in the approach of this study
by varying, for example, the mean of the distribution of the infection rate. 
We assume a single infection incident, alhough there may be 
theoretical ramifications of multiple such events (Pujol et al, 2009). 

 \begin{center}
\begin{table}[!h]
\caption{Variables and parameters}
    \begin{tabular}{lll}
  \hline
  {\er  Symbol}   & {\er Description}  & {\er Units } \\
  \hline
 T & {\er Density of target CD4+ T cells}  & T-cells  $\mu$l$^{-1}$  \\
 T$^*$ & {\er Density of productively infected CD4+ T cells} &  T$^*$-cells $\mu$l$^{-1}$  \\
V & {\er Density of virions}  & Virions $\mu$l$^{-1}$\\
   {\er  $ \lambda$} & {\er Rate of arrival of target CD4+ T cells } &  T-cells $\mu$l$^{-1}$  day$^{-1}$   \\
      {\er  $\mu$} & {\er Per capita rate of decrease of target CD4+ T cells } &  day$^{-1}$   \\
            {\er  $k$} & {\er Rate of conversion of T to T$^*$ by virus } & day$^{-1}$ (virions $\mu$l$^{-1}$)$^{-1}$   \\  
                  {\er  $\delta$} & {\er Per capita rate of decrease of productively infected CD4+ T cells } &  day$^{-1}$      \\
                    {\er  $p$} & {\er Rate at which T$^*$ cells produce virus} &  virions T$^*$cells$^{-1}$ day$^{-1}$    \\
                       {\er  $c$} & {\er Per capita rate of decrease of virions } &   day$^{-1}$   \\
  \hline 
   \end{tabular}
\end{table} 
\end{center}

\subsection{Equilibrium analysis}
Equilibrium point analysis of the 
system of differential equations (1)-(3) has been carried out 
by several authors (for example Bonhoeffer et al. (1997),
Stafford et al., 2000, 
Tuckwell and Wan, 2000). There are two equilibrium points, denoted by
$P_1$ and $P_2$.  These occur at 
\be P_1=\bigg( \frac{\lambda}{\mu}, 0, 0 \bigg) \ee
and 
\be P_2=\bigg(  \frac{c \delta}{kp},  \frac{\lambda}{\delta}-  \frac{c \mu}{kp},    \frac{\lambda p}{c\delta}  
- \frac{\mu}{k} \bigg). \ee  
To discuss the outcomes for an infection by HIV, define
\be R= \frac{c \delta \mu}{k \lambda p} \ee
and note the following possibilities.

\subsubsection{Case 1, $R > 1$.}
It is clear that in this case 
the equilibrium values of $T^*$ and $V$ are negative
so that $P_2$ is outside the first octant at unbiological values.
This might occur, for example, if the if the arrival rate  of target
 CD4 + T cells from the thymus is sufficiently small to make 
\be \lambda  <  \frac{c \mu \delta}{kp}. \ee
Alternatively,  large enough values of one or more of $c$, $\mu$ and $\delta$ and/or
small enough values of one or both of $k$ and $p$, will also tend to make this inequality
hold. 
Effectively then there is just one critical point $P_1$ which is at zero virions ($V=0$) and zero
productively infected cells ($T^*=0$) with the unperturbed equilibrium value
of target CD4+ T cells $T=\frac{\lambda}{\mu}$. Thus
\be R > 1  \implies \lim_{t \rightarrow \infty}  V(t) \rightarrow 0. \ee  
This means that, according to the model, the virus goes extinct and the infected host
is cleared of the HIV virus.   If the host immune system parameters do not change
as a consequence of the infection, then a second dose of virions would meet also
with extinction and this process could, theoretically,  be repeated indefinitely.

\subsubsection{Case 2, $R < 1$}
In this case both critical points $P_1$ and $P_2$ are at biologically meaningful values
in the first octant.  The point $P_1$  is an unstable saddle point 
and the point $P_2$ is an asymptotically stable spiral point. 
Thus when case 2 conditions are fulfilled,
\be \lim_{t \rightarrow \infty} V(t) = V_f >0, \ee
where 
\be V_f =  \frac{\lambda p}{c\delta}  
- \frac{\mu}{k}  \ee  
 is the equilibrium value of the number density of virions. 

Note that in Tuckwell and Wan (2000) the possibility of an extra term
$-kTV$ in (3) was considered leading to the replacement of $p$ by $p-\delta$
in the expression for $R$. 
However, since $\delta<< p$, the additional term, which was taken into account in
the calculations presented below, makes very little difference in determining
the probabilities which we shall calculate. Note that at $R = 0$, the bifurcation point,
 there is only one critical point $P_1$ so that $V \rightarrow 0$  as in Case 1
 (Tuckwell and Wan, 2000). 

It is apparent that factors which make $R$ smaller, promote the persistence
of the diseased state and factors which make $R$ larger inhibit the host viral
population. From the definition of $R$ it is clear that larger values of the
following promote the persistence of HIV infection: $k$, leading to greater fre-
quency of virus-T-cell interactions; $\lambda$, giving a larger density of target T-cells;
and $p$, the number of virions produced per T*-cell. Similarly, larger values of
the following tend to inhibit the HIV population: $c$, the viral clearance rate;
$\delta$, the rate of disappearance of T*-cells; and $\mu$, the rate of disappearance of
target T-cells. We use the term {\it promoters} for $k$, $\lambda$  and $p$ and we call $c$,
$\delta$ and $\mu$,  {\it inhibitors}, 
 these two groups being treated differently in the analysis
below.

\section{Methods}
Let us denote the random variables representing the parameters by the
symbols given in Table 2. 
\begin{center}
\begin{table}[h]
    \caption{Notation for random parameters}
\begin{center}
\begin{tabular}{|c|c|}
\hline
Parameter & Random \\
\hline
$\lambda$ & $\Lambda$  \\
$ \mu$ & M   \\
$k$ & $K$  \\
$p$ & P  \\
$\delta$ & $\Delta$  \\
\hline
\end{tabular}
\end{center}
\end{table}
\end{center}
Whether case 1 or case 2 applies depends on the values of the set of 6 random
variables
\be U= \{ \Lambda, M, K, \Delta, P, C\}.  \ee
The calculation which we address in this article is the determination of the
probability of occurrence of the various values of ${U}$.
In particular we will attempt to estimate
\be p_E= {\rm Prob}  \{  U \in E \}  \ee
where $E$ is the set of values of the 6 parameters which lead to $P_1$  being an
asymptotically stable node. This will provide an estimate of the
probability that the virus goes extinct, even in the absence of any drug treatment.
Note that this is a population probability describing the chance that the virus
does not persist in a 
randomly selected member of the population of hosts, not a probability that in a given
individual the virus will go extinct. That is, if the host population size is $n$ 
and $N_R$ is the number who recover from the viral infection without
treatment\, then 
\be  {\rm E}(N_R)=Np_E.\ee

\subsection{Estimation of $p_E$}
All the parameters in $U$ are non-negative and continuous so that
they may be ascribed probability density functions.  Let us denote
the densities  of the six components by
\be \label{den} f_{\Lambda}, f_{M}, f_{K}, f_{\Delta}, f_{P}, f_C. \ee
It is feasible to determine $p_E$ analytically if certain simplifying
assumptions are made about the distributions of the various parameters. 
However, a better approach is to be guided by empirical evidence 
to estimate the various densities $f$  and then use simulation to find
$p_E$.  

Stafford, \textit{et. al.} (2000) gave estimates (their Table 2)
for the values of the parameters
 $\mu, k, \delta$ and $p$ for 10 patients infected by HIV. 
 In Table 3, we give 
the means, standard deviations, and upper and lower 95\% confidence limits
for the means as well as the maxima and minima for these four parameters.
As in Stafford et al. it is assumed that $c = 3$ for all patients and also that
$\lambda=10\mu$. The latter is consistent with  Stafford et al.'s values for
$R_0$ (which is the reciprocal of $R$ defined in (6)). However, different values of  
$\lambda$ are obtained
using Stafford et al.'s equilibrium value for $T = T_{ss}$. Thus using  $\lambda= 10\mu$, the
mean is $\ov{\lambda} = 0.1089$ whereas using the values of $T_{ss}$
 one obtains  $\ov{\lambda} = 0.193$. 
In the simulations below we use the values  $\lambda=10\mu$.

\begin{center}
\begin{table}[!h]
\caption{Some statistics for 10 patients, from Stafford et al. (2000)}
\begin{center}
\begin{tabular}{|c|c|c|c|c|}
\hline
 & $\mu$  &  $k$ & $\delta$ & $p$\\
\hline
Mean & 0.01089 &  $1.179 \times 10^{-3}$ & 0.3660 & 1426.8 \\
\hline
Standard deviation & 0.005727 &  $1.422 \times 10^{-3}$ & 0.193 & 2049.36  \\
\hline
Minimum & 0.0043  & 0.19 $\times 10^{-3}$& 0.13 & 98\\
\hline
Maximum  & 0.020 & 4.80 $\times 10^{-3}$& 0.80 & 7100 \\
\hline
Upper 95\% conf limit  & 0.00734 & $0.2976 \times 10^{-3}$ & 0.246377 & 156  \\
\hline
Lower 95\% conf limit & 0.0144396 &  $2.560 \times 10^{-3}$ & 0.4856 & 2697 \\
\hline
\end{tabular}
\end{center}
\end{table}
\end{center}


  We note that based
on the Stafford et al. (2000) data, a 95\% confidence interval for the mean
of $R$ is (0.0710, 0.1409) and so the chances of $ V_f = 0$ is practically zero, as
would be expected from a group of patients who definitely  have a sustained
HIV infection.
The same conclusion arises from the 95\% confidence interval
(4.089, 7.310) for the mean value of $R_0$.

\subsection{Distributions of parameters}

For the estimation of the probability that the virus goes extinct after infect-
ing a new host, the distributions of the random variables 
 $ \Lambda, M, K,  \Delta, P, C$  are
required. These are not known with certainty so we assume that, by a central-
limit theorem argument, the parameters wil be approximately normally dis-
tributed, although lognormal distributions were previously employed (Ciupe
et al., 2006). For the density of $C$  we take a delta-function concentrated
at $c = 3$. If we take the remaining variables  $ \Lambda, M, K,  \Delta, P$
 to have normal
distributions then there is always going to be a small probability mass for
each variable at values less than zero, which is biologicaly unrealistic for the
present set of parameters. Hence we have chosen truncated normal distri-
butions. For a normal random variable $X$ with mean $m$, standard deviation
$\sigma$, truncated to be on the interval $(\alpha, \beta)$, the probability
density function is (Johnson and Kotz, 1970)
\be     f_X(x)=      \frac { \phi\big(    \frac {x-m}{\sigma}  \big)     }  
  { \sigma  \big[ \Phi\big(    \frac {\beta-m}{\sigma}  \big) -  \Phi\big(
    \frac {\alpha-m}{\sigma}  \big) \big]  }, \alpha < x < \beta,   \ee 
where $\phi$ and $\Phi$ are the density and distribution function
for a standard normal random variable.  The mean of the
truncated variable is
\be    E[X]=  m +   \frac { \phi\big(    \frac {\alpha-m}{\sigma}  \big)    - \phi\big(    \frac {\beta-m}{\sigma}  \big)  }  
  {   \big[ \Phi\big(    \frac {\beta-m}{\sigma}  \big) -  \Phi\big(
    \frac {\alpha-m}{\sigma}  \big) \big]  } \sigma  \ee 
which is useful for comparing with the mean of the parent distribution. 

\subsection{Sampling procedures}

In order to estimate $p_E$ we generate samples of size 10,000 for each of the 
parameters $\mu, \lambda, k, \delta, p$ using the truncated normal probability density functions. The
distributions are specifed in terms of the four quantites: mean, standard 
deviation (both chosen before truncation), and the lower and upper truncation
points. For the upper and lower bounds, three approaches were used.

{\it Method 1}: Use the means and standard deviations of the Stafford et al.
(2000) data of Table 3 and use the minimum value of each parameter as $\alpha$
and the maximum value as $\beta$  for the corresponding random variable.

{\it Method 2}: Take  $\alpha$ to be half of the minimum value of the Stafford et al. (2000) data,
and $\beta$ as twice the maximum value of the Stafford et al. (2000) data. The
motivation here is to extend the ranges of the parameter data of Stafford
et al. (2000), because those data are for patients who became infected and
remained infected, whereas some members of the population may become
infected with HIV and recover spontaneously if their immune system is able
to eliminate the virus as  in the case of  $R > 1$. Again
we use the original standard deviation given in Table 3. However, instead
of the actual mean, in order to examine outcomes with more extreme but
possible parameter values, we may also use the estimated lower and upper
confidence limits of the means. 

{\it Method 3}:
Here the means of the parameters which are promoters, $k, \lambda, p$ 
are multiplied by 0.9, whereas the means of the parameters $\delta$ and $\mu$
 which are 
inhibitors are multiplied by 1.1. The same standard deviations as in Method
1 are employed and the 95\% confidence limits for the means are recalculated.
We do this using the same sets of lower and upper bounds for the parameter
densities as for Method 2.

{\it Method 4}: This method is also motivated by the shifting of the parameter 
distributions of the Stafford et al. (2000) to account for bias in the data due
to the fact that they were taken from patients in whom HIV persisted. Here
we multiply the means and the lower and upper bounds (taken to be the
minimum and maximum values of the parameters in Table 3, as in Method
1) of the promoter parameters $k, \lambda, p$  by values $s_p < 1$, and we multiply
the means and the upper and lower bounds of the inhibitor parameters  $\delta$ and $\mu$ 
 by values $s_d > 1$. Such choices must lead to a greater chance of extinction
of the virus.
In all cases the value of $R$, which is now a random variable, is computed, and hence the
 probability $p_E$ that $P_1$ is an asymptotically stable node
or equivalently that extinction of the host HIV population occurs, can be
estimated. 

\section{Results}
For Method 1, the use of the means and standard deviations of the data in
Table 3 as well as the minima and maxima as the truncation limits resulted
in the following means from formula (16) for the parameters: (with original
means in brackets)
$ E[M] = 0.0114(0:01089), E[K] = 1.6401 \times 10^{-3}, (1.179  \times 10^{-3}),
 E[\Delta] = 0.3965(0.3660)$ and  $E[P] = 2141.1 (1426.8)$.  Thus, the truncation
procedure results in increases in the means for all of these 4 parameters. 

\subsection{Calculated extinction probabilities}
In total, as described above for Methods 1,2 and 3,  there were 
many different ways in which
 the parameters of the distributions of the randomized
parameters were chosen. To illustrate, there are shown in Figure 1,
the histograms for the parameter $\lambda$ with truncated 
distributions according to Methods 1-3.

\centerline{\includegraphics[width=5.1in]{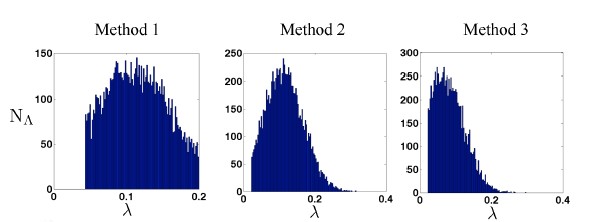}}

\begin{figure}[h]
\caption{The truncated normal distributions of $\Lambda$  according to 
the densities described in the text, with means chosen to be the 
values of Table 3.}
\label{}
\end{figure}

 For each Method and for each
choice of parameterization of the distributions the calculated
results for the probabilities of extinction $p_E$  were as follows (Table 4). 

\begin{center}
\begin{table}[h]
    \caption{Estimated extinction probabilities, $p_E$}
\begin{center}
\begin{tabular}{|c|c|c|c|}

\hline
&   Method 1 & Method 2  & Method 3 \\
\hline
Maximum value & 0.0685 &  0.1354 & 0.1376 \\
\hline
Minimum value  & 0.0063 & 0.0066 & 0.0158 \\
\hline
Average value & 0.0248 & 0.0450 & 0.0693 \\
\hline
\end{tabular}
\end{center}
\end{table}
\end{center}

Figures 2 and 3 show geometrically, by means of three-dimensional
scatter diagrams,  the situation with regard to the critical 
points in two extreme cases. If $P_2$ falls outside the first octant then
spontaneous recovery can, according to the model predictions, occur.
In Figure 2 is shown the spatial distribution of  $P_1$ and $P_2$ values when the
probability of recovery is small, with most $P_2$-values falling in the first octant. 
In contrast, Figure 3 shows a case where the probability of recovery is
much higher as many more $P_2$-values fall outside of the first octant.
The sample sizes used for these figures were reduced to 2000 from the
usual 10000 to make the
figure files manageable. 

\centerline{\includegraphics[width=3.5in]{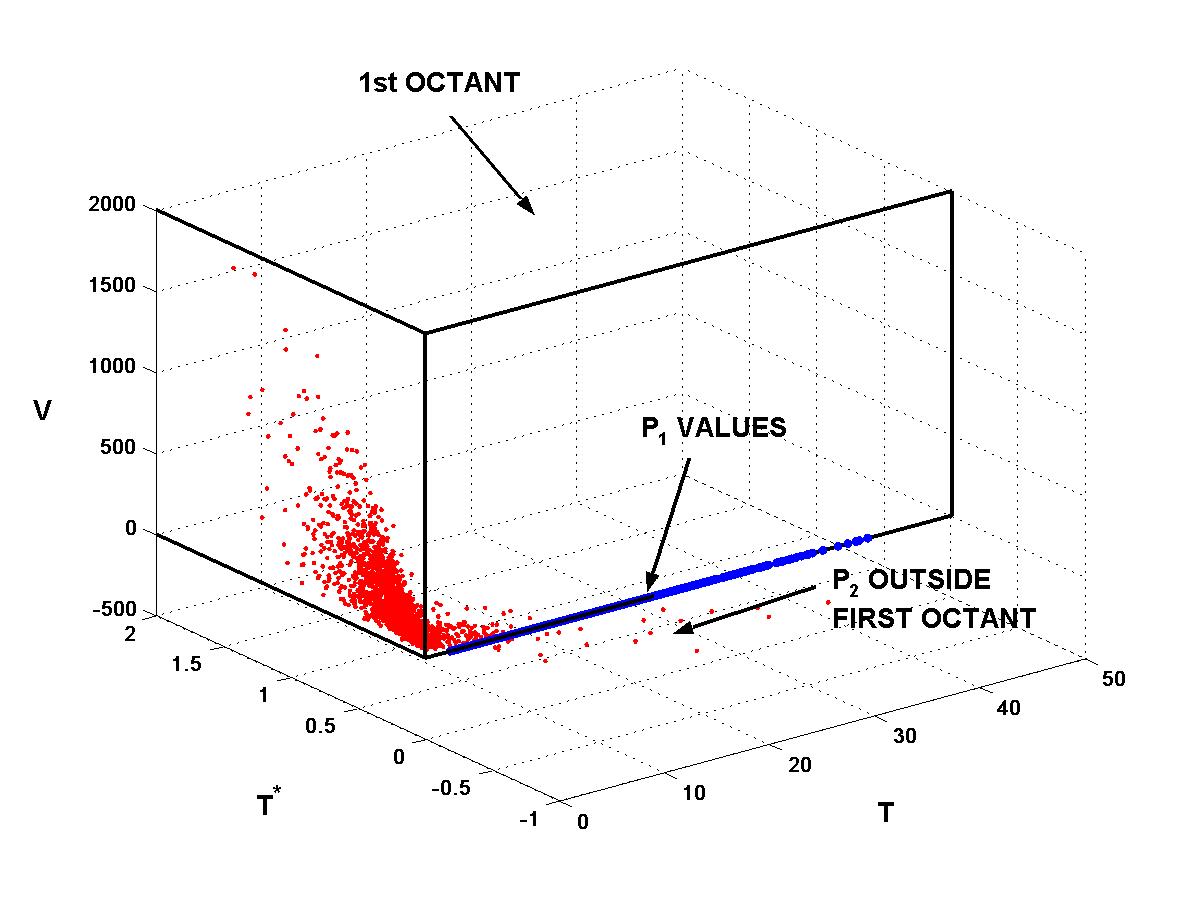}}

\begin{figure}[h]
\caption{A sample distribution of the positions in $(T, T*, V)$-space of the
two critical points $P_1$ (blue circles) and $P_2$ (red circles) when the probability
of recovery from HIV infection is very small. Only a relatively small number
(about 1.5\%)
of points $P_2$ lie outside the first octant. Here the sample is generated
by Method 1 using the actual minima, mean and maxima as given in Table 3
for the parameters of Stafford et al. (2000) for use in the truncated normals.
Sample size 2000.}
\label{}
\end{figure}

We also explore calculations in which we shift the distributions to
 account for bias in the Stafford data towards
 patients in which the virus persisted, which we have called Method 4.
 Results are given in Table 5 and see also Figure 3. 

\begin{center}
\begin{table}
    \caption{Values of $p_E$,   Method 4.}
\begin{center}
\begin{tabular}{|c|c|c|}

\hline
$s_d$ & $s_p$ & $p_E$ \\
\hline
1.1 & 0.9 & 0.9747 \\
1.2 & 0.8 & 0.9499 \\
1.25 & 0.75 & 0.9329 \\
1.3 & 0.7 & 0.9160 \\
1.4 & 0.6 & 0.8586 \\
1.5 & 0.5 & 0.7601 \\
\hline
\end{tabular}
\end{center}
\end{table}
\end{center}

\centerline{\includegraphics[width=3.5in]{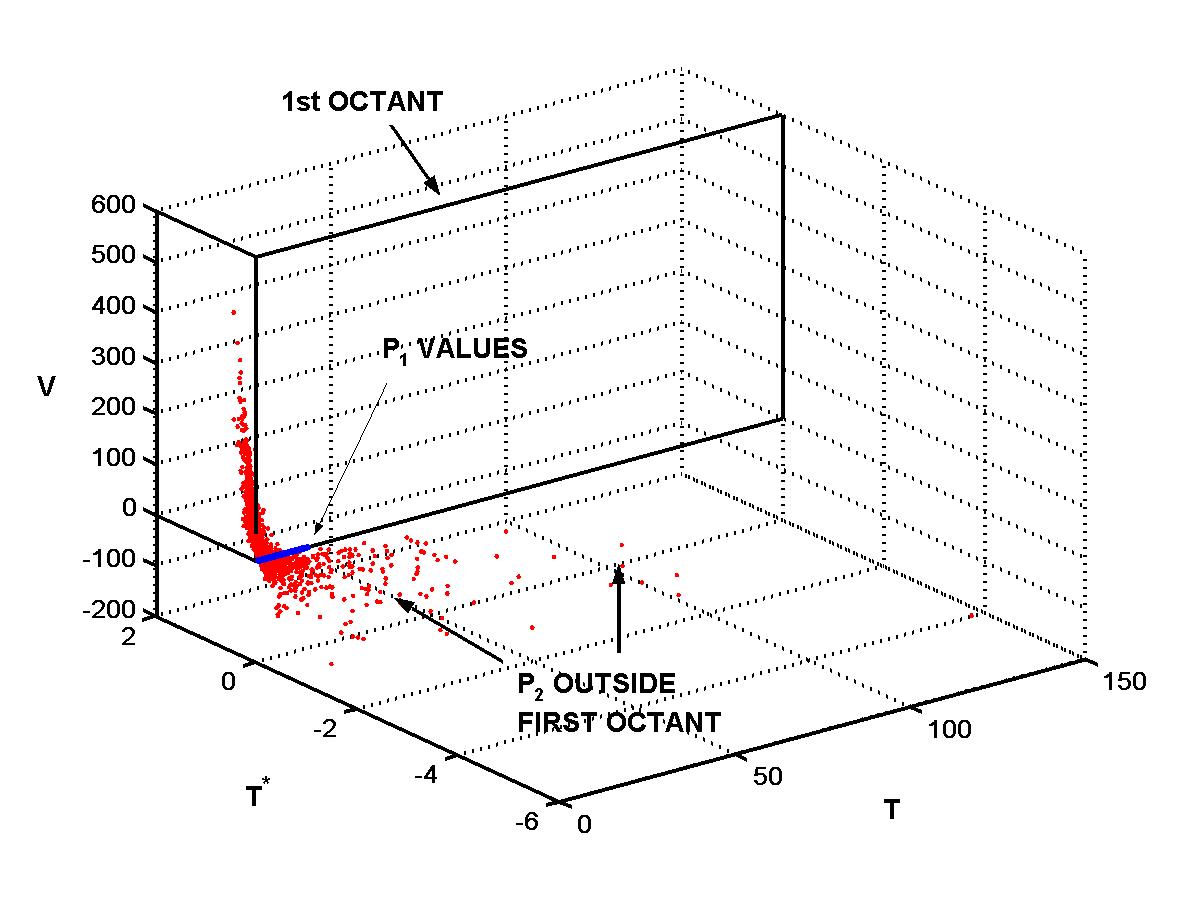}}
\begin{figure}[h]
\caption{A sample distribution of the positions in $(T, T*, V)$-space of the
two critical points $P_1$ (blue circles) and $P_2$ (red circles) when the probability
of recovery from HIV infection is relatively large.
 About 25\% of
the points $P_2$ lie outside the first octant. Here the sample is generated using
Method 4 where the means and upper and lower truncation points for the
distributions for $k, \lambda, p$  are multiplied by 0.5 and those of $\mu$ and $\delta$
 are multiplied
by 1.5.
Sample size 2000.}
\label{}
\end{figure}

\section{Discussion}
The probabilities $p_E$ determined above represent the
chance that the parameters of the virus-immune system dynamics
are such that the only equilibrium is at zero virions and hence the
virus would, theoretically, be eliminated without medical intervention.
The values of $p_E$ given in Table 4 have approximate upper bounds of
0.07 for Method 1 and  0.14 for Methods 2 and 3. These probabilities
are quite small, which is due, at least in part, to the fact that
the estimated parameter distributions are biased, being derived from a sample
of individuals, all of whom were infected by HIV for tracked 
periods of from 46 days to several hundred days.  A histogram
of steady state $V=V_f$ values for Method 1 is shown in Figure 4.
Here a small fraction of values is less than or equal to zero,
and ones just greater than zero could be driven to zero with noise.

\centerline{\includegraphics[width=3.5in]{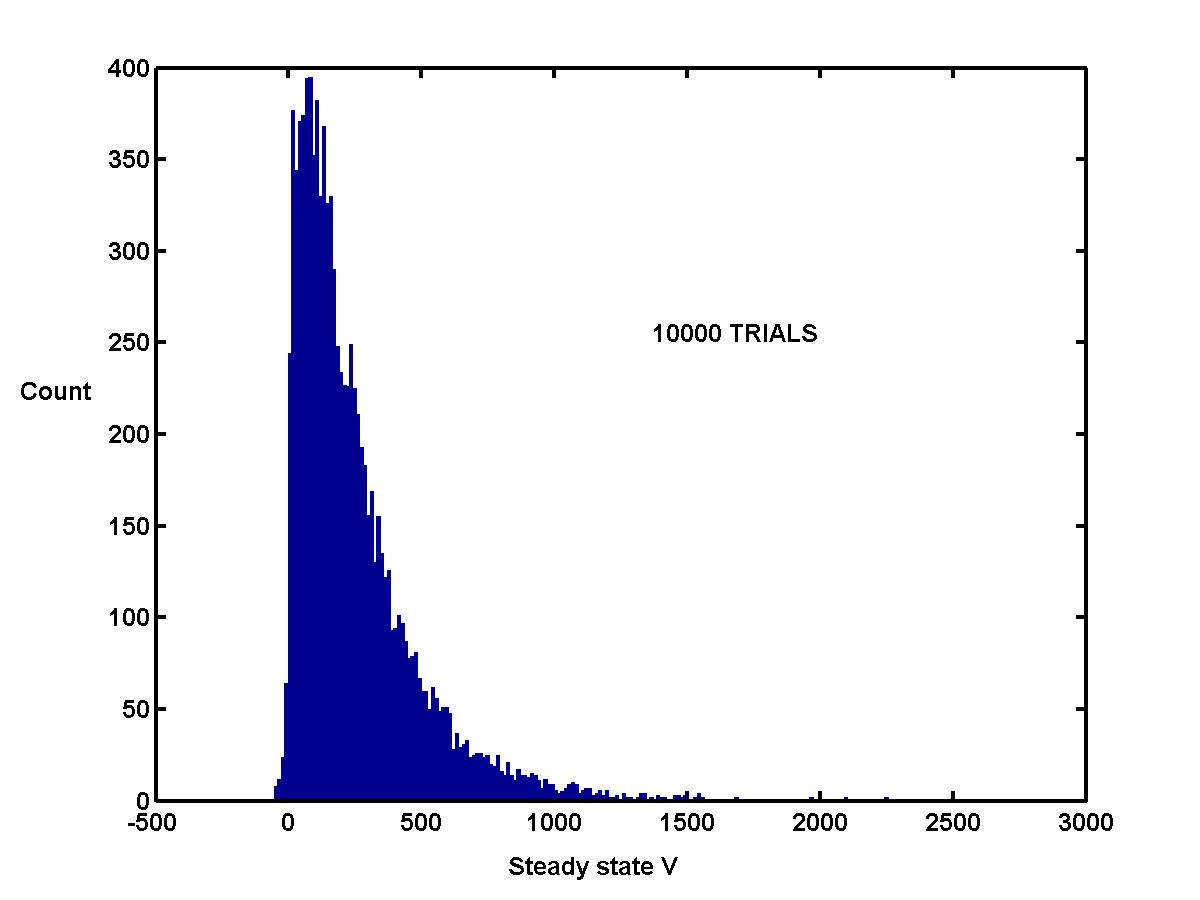}}

\begin{figure}[h]
\caption{Histogram of steady state $V$ values when the parameter
distributions are obtained by the conservative method 1.}
\label{}
\end{figure}

When the ranges of the parameters were extended
by not extreme amounts in an attempt to counter 
the bias towards an infected group,  values of $p_E$ were obtained as high as about 0.24. 
Thus it is feasible that in the general population, within the framework 
of the standard 3-component model, that a considerable
percentage of individuals could be infected by HIV but the
virus would subsequently be eliminated
without intervention which would constitute spontaneous recovery.

It is noted that the virus population in patient 3 of the Stafford
et al. data
had dwindled to 0.1\% of its maximum after 50 days. 
Furthermore, if one takes the values of the parameters
for all 10 patients which lead to the largest value of $R=R_{max}$,
which are $\delta=0.8$, $k=0.00019$, $p=98$ and the standard
values $c=3$ and $\mu/\lambda=0.1$, 
then $R_{max}=12.9$ which far exceeds the
value required to make the only equilibrium point
the one at zero virions.

The probability of transmission of HIV has been estimated in various
groups. Gray et al. (2001) found an average value of 0.0011  per coital
act in heterosexual couples in Rakai, Uganda and Wawer
et al. (2005) found a value of 0.0082 for a similar population.
See also Chakraborty et al. (2001).
Higher values, around 0.031, had been obtained for a group of male
military trainees who interacted with female prostitutes in
Thailand (Mastro et al., 1994).  In the Ugandan studies, two of the
main factors influencing transmission were time since infection
of the index partner and degree of ulceration.

  Many findings
in such articles, and in Downs et al. (1996), concern transmission in stable monogamous
discordant couples, in which one partner is HIV-positive and yet,
despite frequent and long-term possibilities of transmission,
the HIV-negative partner does not seroconvert, even 
in the absence of condom use.  This was also reported
in the metaanalytical study of Powers et al. (2008), where
there was found to be no transmission even after hundreds
of contacts between members of discordant couples. Similarly, 
 transmission had been remarkably 
found to not occur in some
 individuals despite multiple high-risk
 sexual exposures (Rowland-Jones, 1995; Paxton et al., 1996).
 In the latter study,  CD8+ lymphocytes were found to 
 have greater anti-HIV-1 activity than those from nonexposed controls.
There is always the possibility that the non-transmission of the virus in
discordant couples 
is due to properties of the infected partner, such as a low
viral load. Nevertheless, it is feasible, though considered unlikely,  
by HIV experimentalists
and theorists, that in some of these long-term discordant couples, 
the non-index partner does become infected briefly, possibly repetitively, 
but  the virus is subsequently eliminated as predicted by the standard
model. The period(s) of infection
could be brief and the viral load small so that the partner 
remained asymptomatic.

 According to Haynes et al. (1996) 
the existence of individuals who have been exposed
multiple times to HIV and are persistently
HIV-seronegative raises the possibility
that a small percentage of them may be resistant
to HIV, or may have been able to clear the
infection.
 The situation is made less clear 
by virtue of the issues of viral loads below threshold for detection and
other 
false negatives. 
Also pertinent are the rates of HIV infection in children 
born to HIV-infected parents. According to an analysis of
 Rowland-Jones et 
al. (1993), 60-85\% of children exposed either before or after
birth to HIV were not infected.  In a European study
(N\"ostlinger et al., 2004), a study of 165 HIV-affected
families with 279 children, found that 68\% of the 
children were HIV-negative.

Much has been written about host immune properties or
properties of the virus which
may lead to the fending off of HIV infection (Haynes et al., 
 1996; Lo Caputo et al., 2003)
or as in the 
case of elite controllers (sometimes called long term non progressors
or just HIV controllers) who constitute about
5-10\% of cases, maintaining low viral burdens
and not converting to the AIDS regime.  Genetic factors such as
the gene encoding CCL3L1 have been shown to
affect susceptibility to AIDS (Gonzalez et al., 2005)
and certain mutations in the gene encoding the protein CCR5
afford immunity to HIV in mice (Holt et al, 2010). 

There are evidently, however, no documented cases
of clearance of HIV from an an individual with an established
infection (Alan Perelson and Marc Pellegrini, personal
communications). There were cases  reported in the press of changes
in individuals 
from HIV-positive to HIV-negative,
the most noteworthy being that of Alan Stimpson. 
The standard model employed here predicts, under
reasonable assumptions on the distributions of parameters,
that a small percentage of the population might be able to
clear HIV after infection. As pointed out above, it would
be hard to verify or refute this prediction
 if the duration of infection and the
viral load were both small.   In actuality, there are
several immune system components omitted in the standard
3-component model, such as latently-infected cells, 
cytotoxic T-cells (CD8+ T-cells, CTL),  natural killer cells  and 
dendritic cells. The incorporation of these
additional components would make the mathematical
modeling much more complicated but may yield 
more insight into the possibility of an individual's
clearing an HIV infection.

\section{\bf Acknowledgements} 
We thank Dr Alan Perelson of Theoretical Biology \& Biophysics, Los Alamos 
National Laboratory, Los Alamos, New Mexico, USA,  and 
 Dr Marc Pellegrini of the Walter and Eliza Hall Institute
of Medical Research, Parkville, Victoria, Australia, for 
useful correspondence.

\nh Bonhoeffer N Coffin JM and Nowak MA (1997). 
Human Immunodefciency virus drug therapy and virus
load. J Virology 71, 3275-3278.

\nh Chakraborty H et al. (2001)
Viral burden in genital secretions determines male-to-female
sexual transmission of HIV-1: a probabilistic
empiric model. AIDS 15: 621-627.

\nh Ciupe MS et al. (2006) Estimating kinetic parameters from 
HIV primary infection
data through the eyes of three different mathematical models.
Math Biosci 200: 1-27.

\nh Downs AM et al. (1996) Probability of heterosexual transmission of HIV: 
relationship to the number of unprotected sexual contacts. JAIDS 11: 388-395.

\nh Egger M et al. (2002) Prognosis of HIV-1-infected patients starting highly active
antiretroviral therapy: a collaborative analysis of prospective studies.
The Lancet  360,  119-129.

\nh Essunger, P and  Perelson, A.S. (1994).  Modeling HIV infection of CD4+ T-cell
subpopulations.  J. Theor. Biol. 170: 367-391

\nh Finzi D, Siliciano RF (1998)
Viral Dynamics in HIV-1 Infection. Cell 93, 665-671.

\nh Gallant JE et al. (2006) Tenofovir DF, emtricitabine, and efavirenz vs.
zidovudine, lamivudine, and efavirenz for HIV.  NEJM 354, 251-260. 

\nh Gonzalez E et al. (2005) The influence of CCL3L1 
gene-containing segmental duplications
on HIV-1/AIDS susceptibility. Science 307: 1434-1440. 

\nh Gray RH et al. (2001) 
Probability of HIV-1 transmission per coital act in monogamous,
heterosexual, HIV-1-discordant couples in Rakai, Uganda. 
 Lancet 357: 1149-1153.

\nh Holt N et al. (2010) Human hematopoietic stem/progenitor cells modified
 by zinc-finger nucleases targeted to CCR5 control HIV-1 in vivo.
    Nat Biotech  28: 839-847.

\nh Johnson N, Kotz  S (1970) Distributions in Statistics: Continuous Univariate
Distributions 1.  Houghton Mifflin, New York.

   \nh   Kamina A, Makuch RW, Zhao H  (2001)  A stochastic modeling of early HIV-1 population
          dynamics. Math Biosci 170: 187-198.

          \nh Kim H, Perelson AS (2006) Dynamic characteristics of HIV-1 reservoirs. Curr Opin
HIV AIDS 1, 152-156.

       \nh Lo Caputo S et al. (2003)
Mucosal and systemic HIV-1-specific immunity in HIV-1-exposed but 
uninfected heterosexual men. AIDS 17: 531-539. 

\nh Mastro TD et al. (1994) Probability of female-to-male transmission of
 HIV-1 in Thailand. Lancet 343: 204-207.

          \nh   Merrill SJ (2005)  The stochastic dance of early HIV infection.
          J. Comp Appl Math 272: 74-79. 
          
\nh N\"ostlinger C et al. (2004) 
Families affected by HIV: parents' and children's characteristics
 and disclosure to the children. AIDS Care 16: 641-648.

          \nh Nowak MA, Bangham CRM (1996). Population
dynamics of immune responses to persistent viruses.
Science 272, 74-79. 
          
  \nh Paxton WA et al. (1996)   
Relative resistance to HIV−1 infection of CD4 lymphocytes
 from persons who remain uninfected despite multiple high-risk sexual exposures. 
Nat Med  2: 412-417.

\nh Pearson JE et al. (2010) Stochastic theory of early viral infection: continuous
versus burst production of virions. PLoS Comp Biol Volume 7: e1001058.

\nh Perelson AS (2002) Modelling viral and immune system dynamics. Nat Rev Immunol 2, 28-36.  

\nh  Perelson  AS, Kirschner DE, De Boer R (1993)
Dynamics of HIV infection of CD4 T cells. Math Biosci
114: 81-125.

 \nh  Perelson AS, Neumann AU, Markowitz M et al. (1996) HIV-1 dynamics in vivo: 
 virion clearance rate, infected 
 cell life-span, and viral generation time.  Science 271: 1582-1586.

 \nh   Phillips AN (1996)  Reduction of HIV concentration during acute infection: 
idependence from a  specific immune response. Science 271: 497-499.

\nh Phillips AN et al. (2001) HIV viral load response to
antiretroviral therapy according to the
baseline CD4 cell count and viral load. JAMA 286: 2560-2567.

\nh Pierson T,  McArthur J, Siliciano RF (2000) Reservoirs for HIV-1: mechanisms for viral
persistence in the presence of antiviral immune
responses and antiretroviral therapy. Ann Rev Immunol 18: 665-708.  

\nh  Pope M, Haase AT (2003) Transmission, acute HIV-1 infection and the quest for
strategies to prevent infection. Nat Med 9:  847-852.

\nh Powers KA et al. (2008) Rethinking the heterosexual infectivity
 of HIV-1: a systematic review and meta-analysis.  
Lancet Infect Dis 8: 553-563.

\nh  Pujol JM et al.  (2009)
The effect of ongoing exposure dynamics in dose
response relationships.  PLoS Comp Biol 5:  e1000399. 

\nh Rowland-Jones S et al. (1993)
HIV-specific cytotoxic T-cell activity in an HIV-exposed but
 uninfected infant. Lancet 341: 860-861. 

\nh Rowland-Jones S et al. (1995) HIV-specific cytotoxic T-cells in HIV-exposed
but uninfected Gambian women. Nat Med 1: 59-64.  

 \nh   Stafford MA et al.  (2000)
            Modeling plasma 
            virus concentration during primary HIV infection. 
J Theor Biol 203: 285-301.

\nh  Tan WY, Wu H (1998)  Stochastic modeling of the
 dynamics of CD4+ T-cells infection 
by HIV and
some Monte-Carlo studies.  Math Biosci 147: 173-205.

\nh   Tuckwell HC,  Le Corfec E (1998)   A stochastic model for
 early HIV-1 population
dynamics.  J Theor Biol 195: 451-463.

 \nh Tuckwell HC, Shipman PD, Perelson  AS (2008)  
 The probability of HIV infection in a
new host and its reduction with microbicides. Math  
Biosci 214: 81-86.

\nh Tuckwell HC, Wan FYM (2000)  Nature of equilibria and effects of
 drug treatments in some
simple viral population dynamical models.
IMA J Math Appl Med Biol 17: 311-327.

  \nh     Vaidya RM et al.  (2010)
Viral dynamics during primary simian immunodeficiency virus
infection: effect of time-dependent virus infectivity. J Virol 84: 4302-4310.

\nh Wawer MJ et al. (2005) Rates of HIV-1 transmission per coital act,
by stage of HIV-1 infection, in Rakai, Uganda.
 J Infect Dis 191: 1403-1409.

 \end{document}